\documentclass[aps,prl,twocolumn,showpacs,10pt,superscriptaddress,preprintnumbers,footinbib]{revtex4-1}
\usepackage{amsmath}
\usepackage{physics}
\usepackage{graphicx}
\usepackage{amssymb}
\usepackage[colorlinks,citecolor=red]{hyperref}

\usepackage{color}    
\usepackage{ulem}
\usepackage{multirow}

\usepackage{ulem}

\newcommand{\bqa}{\begin{eqnarray}}
\newcommand{\eqa}{\end{eqnarray}}
\newcommand{\beq}{\begin{equation}}
\newcommand{\eeq}{\end{equation}}

\newcommand{\nn}{\nonumber}
\allowdisplaybreaks

\begin{document}

\title{Analytic three-loop QCD corrections to top-quark  and  semileptonic $b\to u$ decays}

\author{Long-Bin Chen}
\affiliation{School of Physics and Materials Science, Guangzhou University, Guangzhou 510006, China}

\author{Hai Tao Li}
\email{haitao.li@sdu.edu.cn}
\affiliation{School of Physics, Shandong University, Jinan, Shandong 250100, China}

\author{Zhao Li}
\email{zhaoli@ihep.ac.cn}
\affiliation{Institute of High Energy Physics, Chinese Academy of Sciences, Beijing 100049, China}
\affiliation{School of Physics Sciences, University of Chinese Academy of Sciences, Beijing 100049, China}
\affiliation{Center for High Energy Physics, Peking University, Beijing 100871, China}

\author{Jian Wang}
\email{j.wang@sdu.edu.cn}
\affiliation{School of Physics, Shandong University, Jinan, Shandong 250100, China}

\author{Yefan Wang}
\email{wangyefan@sdu.edu.cn}
\affiliation{School of Physics, Shandong University, Jinan, Shandong 250100, China}

\author{Quan-feng Wu}
\email{wuquanfeng@ihep.ac.cn}
\affiliation{Institute of High Energy Physics, Chinese Academy of Sciences, Beijing 100049, China}
\affiliation{School of Physics Sciences, University of Chinese Academy of Sciences, Beijing 100049, China}


\begin{abstract}
We present the first analytic results of N$^3$LO QCD corrections to the top-quark decay width. We focus on the dominant leading color contribution, which includes light-quark loops. At NNLO, this dominant contribution accounts for 95\% of the total correction.
By utilizing the optical theorem, the N$^3$LO corrections are related to the imaginary parts of the four-loop self-energy Feynman diagrams,
which are calculated with differential equations. The results are expressed in terms of harmonic polylogarithms, enabling fast and accurate evaluation. 
The third-order QCD corrections decrease the LO decay width by 0.667\%, and the scale uncertainty is reduced by half compared to the NNLO result.
The most precise prediction for the top-quark width is now 1.321 GeV for $m_t=172.69$ GeV.
Additionally, we obtain the third-order QCD corrections to the dilepton invariant mass spectrum and decay width in the semileptonic $b\to u$ transition.
\end{abstract}

\maketitle

The top quark, which is the heaviest elementary particle, has been discovered for about twenty-eight years.
Its mass has been measured to be $m_t=174.41\pm 0.39~(\rm stat.)\pm 0.66~(\rm syst.)\pm 0.25~(\rm recoil)$ GeV \cite{ATLAS:2022jbw},
and its couplings with other particles have been probed with high precision.
This provides necessary input parameters for the calculation of top-quark production processes, e.g., the top-quark pair productions at the hadron colliders.
In realistic simulations of the top-quark events, the decay has to be taken into account as well. 
Therefore, one of the indispensable input parameters in theoretical prediction is the top-quark decay width, denoted by $\Gamma_t$.
It could be modified in new physics models.
A precise determination of the top-quark decay width can be used as a stringent test of the standard model (SM) and a probe of new physics.

Generally, the top-quark decay width can be measured with direct and indirect approaches.
The direct measurement exploits a profile likelihood fit of the observed data, such as the invariant mass of the lepton and $b$-jet, to the template distributions corresponding to different top-quark decay widths.
The ATLAS collaboration has performed a direct measurement using the top-quark pair events in the dileptonic channel at the 13 TeV LHC corresponding to an integrated luminosity of 139 ${\rm fb}^{-1}$, obtaining a width of $\Gamma_t=1.9\pm 0.5$ GeV \cite{ATLAS:2019onj}.

On the other hand, in indirect measurements, the decay width is extracted from quantities that depend on $\Gamma_t$.
The CMS collaboration has measured the branching ratio $B(t\to Wb)/B(t\to Wq)$ with $q=b,s,d$  using the $t\bar{t}$ events in the dileptonic channel at $\sqrt{s}=8$ TeV.
Combined with the measurement of the $t$-channel single top-quark cross section,  the top-quark decay width is determined as  
$\Gamma_t = 1.36 \pm  0.02~(\rm stat.)^{+0.14}_{-0.11}~(\rm syst.)$ GeV \cite{CMS:2014mxl}.
Following the idea proposed to measure the Higgs boson's width \cite{Caola:2013yja},
the top-quark width can also be derived by measuring both the on-shell and off-shell top-quark productions.
The analyses for the single top and top quark pair productions show that an accuracy of 0.3 GeV can be reached \cite{Giardino:2017hva,Baskakov:2018huw,Herwig:2019obz}.

The $e^+ e^-$ collider provides a good opportunity to determine the top-quark mass and width with high precision because the center-of-mass energy is tunable.
The cross sections near the $t\bar{t}$ threshold are very sensitive to the top-quark mass and width.
Assuming an integrated luminosity of $220~ {\rm fb}^{-1}$, the determination of the top-quark width can be carried out with an accuracy at the $2\%$ level \cite{Horiguchi:2013wra,Li:2022iav,CLICdp:2018esa}.

To meet the requirements of both theory and experiment, much effort has been devoted to improving the predictions for top-quark decay.
The NLO QCD corrections decrease the decay width by about $9\%$ \cite{Jezabek:1988iv,Czarnecki:1990kv,Li:1990qf}.
The NNLO QCD corrections provide a  $2\%$ suppression further~\cite{Gao:2012ja,Brucherseifer:2013iv,Chen:2022wit}. 
The analytical form of the NNLO total width has been studied in the  $w \equiv  m_W^2/m_t^2\to 0$ \cite{Czarnecki:1998qc,Chetyrkin:1999ju,Blokland:2004ye,Blokland:2005vq}  and $w\to 1$  \cite{Czarnecki:2001cz} limit, respectively.
The NNLO polarized decay rates have been calculated in \cite{Czarnecki:2010gb,Czarnecki:2018vwh}.
The dependence of the NNLO result on the renormalization scheme and scales was discussed in \cite{Meng:2022htg}.
Recently, the three-loop color-planar master integrals \cite{Chen:2018fwb} and form factors \cite{Datta:2023otd} of the heavy-to-light quark decays were obtained.
The NLO electroweak (EW) corrections \cite{Denner:1990ns,Eilam:1991iz} and off-shell $W$ boson effects \cite{Jezabek:1988iv} have also been computed,
and their contributions almost cancel each other.

The goal of this work is to provide the first analytic results of ${\rm N^3LO}$ QCD corrections.
This is motivated by the fact that the NNLO corrections are beyond the scale uncertainty band of the NLO corrections using the conventional scale variation, i.e., changing the renormalization scale by a factor of two.
It is interesting to see whether the ${\rm N^3LO}$ QCD corrections lie within the scale uncertainty band of the NNLO corrections.
Moreover, our analytic results present special data about the multiloop Feynman integrals and scattering amplitudes, since top-quark decay is the first physical process with massive colored particles that has been calculated at ${\rm N^3LO}$ analytically without any expansion of the loop integrals.

\begin{figure}
    \centering
    \includegraphics[width=1.0\linewidth]{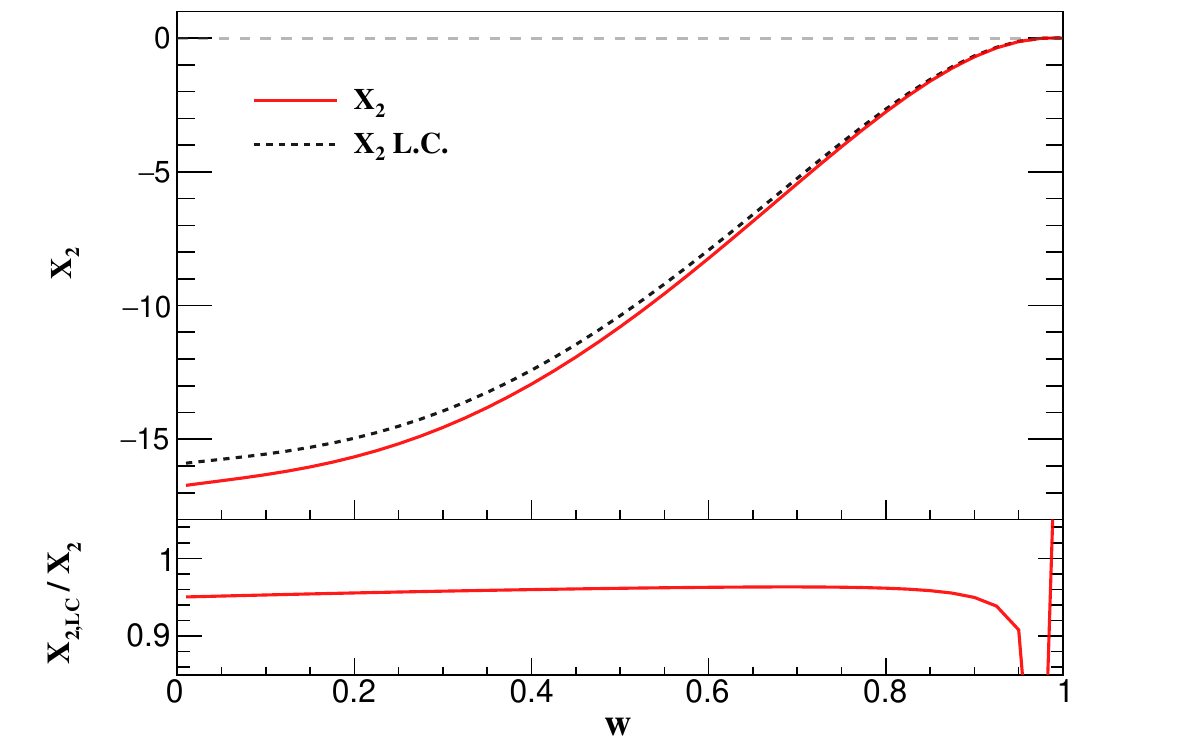}
    \caption{Leading color contribution in $X_{2}$. The upper panel shows the comparison of the full (red line) and the leading color (black dotted line) contribution in $X_2$  as a function of $w$, while the lower panel presents the ratio of the leading color contribution over the full one.    }
    \label{fig:X2LC}
\end{figure}

In the SM, the top quark decays via electroweak interaction to $Wq$ with $q=b,s,d$ at LO,
and the decay rate of $t\to Wq$ is proportional to the square of the Cabibbo-Kobayashi-Maskawa (CKM) matrix element $V_{tq}$.
Since $V_{tb}\approx 0.999$  and $V_{ts}, V_{td}\le 0.04$ \cite{ParticleDataGroup:2022pth},
and the decay rates of $t\to Wq$ differ by a common factor, we consider only $t\to Wb$ in our calculation.
Including higher-order QCD corrections, the top-quark width is written as a series of the strong coupling $\alpha_s$,
\bqa
\Gamma_t=\Gamma_0\left[X_0+\frac{\alpha_s}{\pi}X_1+\left(\frac{\alpha_s}{\pi}\right)^2 X_2+\left(\frac{\alpha_s}{\pi}\right)^3 X_3\right]
\eqa
with  $\Gamma_0=G_F m_t^3|V_{tb}|^2/8\sqrt{2}\pi $.
The first two coefficients $X_0$ and $X_1$ have been calculated analytically thirty years ago \cite{Jezabek:1988iv,Czarnecki:1990kv,Li:1990qf}.
The analytic form of the third coefficient $X_2$ was obtained recently by four of the authors \cite{Chen:2022wit}, and the result is given in different color structures by 
\begin{align}
  X_2 & = C_F \left[C_F X_F + C_A X_A+T_R n_l X_l + T_R n_h X_h   \right] \nonumber\\
  & = C_F\left[ N_c\left(X_A+\frac{X_F}{2}\right) + \frac{n_l}{2}X_l \right.\nonumber\\ 
  & \qquad - \left. \frac{1}{2N_c}X_F + \frac{n_h}{2}X_h  \right]\,,
  \label{eq:x2}
\end{align}
where we have substituted $C_F=(N_c^2-1)/2N_c$, $C_A=N_c$ and $T_R=1/2$ in the second and third lines and $n_l$ ($n_h$) is the number of massless (massive) quark species.
Notice that there is a common color factor $C_F$ for all color structures. 
This is because the process is induced by electroweak interaction between two quarks at LO.
Since this factor persists at higher orders, it is not expanded in $N_c$.
In our case, $N_c=3$, $n_l=5$ and $n_h=1$, and thus we expect that the terms proportional to $N_c$ or $n_l$, denoted as the {\it leading color} contribution, provide the dominant contributions.
As illustrated in Fig.~\ref{fig:X2LC},
the leading color contribution is significantly dominant, 
accounting for around 95\% of the full NNLO 
 correction for $w<0.9$.
Note that  NNLO correction is almost vanishing for $w>0.9$,
and that $w\approx 0.22$  in the top quark decay. 

As in Eq.(\ref{eq:x2}), the result of $X_3$ is decomposed in color structures by 
\begin{align}
  X_3 =\ & C_F\bigg[ N_c^2 Y_A +   \widetilde{Y}_A+\frac{\overline{Y}_A}{N_c^2}   +  n_l n_h Y_{lh} 
   \nonumber \\ &
  + n_l\, \left( N_c  Y_{l} + \frac{\widetilde{Y}_{l}}{N_c} \right)
  +  n_l^2 Y_{l2}  \nonumber\\
  & 
  + n_h\, \left( N_c  Y_{h} + \frac{\widetilde{Y}_{h}}{N_c} \right)
  +  n_h^2 Y_{h2} 
  \bigg]\,.
\label{eq:x3}  
\end{align}
We focus on the leading color coefficients $Y_A$, $Y_{l}$ and $Y_{l2}$ that give the most dominant contributions.
Our goal in this work is to present their analytic form.

In our method, the top-quark decay width $ \Gamma_t$ is related to the imaginary part of the amplitude of the process $t\to Wb\to t$
via the optical theorem, 
\beq
\Gamma_t=\frac{\text{Im}[\mathcal{M}(t\to Wb\to t)]}{m_t}\, .
\eeq
In order to obtain the ${\rm  N^3 LO }$ QCD corrections, the calculation of four-loop self-energy Feynman diagrams is required.

We have used the \texttt{FeAmGen} program \cite{Wu:2023FeAmGen} which employs  \texttt{QGRAF} \cite{Nogueira:1991ex} to generate four-loop Feynman diagrams and amplitudes. 
The amplitudes are manipulated with \texttt{Form} \cite{Kuipers:2012rf} and \texttt{FeynCalc} \cite{Shtabovenko:2020gxv} to perform the Dirac algebra, and expressed as linear combinations of scalar loop integrals.
Then we utilize the C++ version of \texttt{FIRE6} \cite{Smirnov:2019qkx},
which implements the Laporta algorithm \cite{Laporta:2000dsw} in solving the system of the integration-by-parts identities \cite{Tkachov:1981wb,Chetyrkin:1981qh},
to reduce all the scalar integrals into a set of master integrals (MIs). 
All the MIs belong to the integral family defined by 
\bqa
I_{n_1,n_2,\ldots,n_{14}}=\frac{1}{\pi}\textrm{Im} \int\frac{{\mathcal D}^d k_1~{\mathcal D}^d k_2~{\mathcal D}^d k_3~{\mathcal D}^d k_4}{D_1^{n_1}~D_2^{n_2}~D_3^{n_3}~\cdots~ ~D_{14}^{n_{14}}}
\label{eq:def}
\eqa
with the propagators 
\begin{align}
D_1 & =(k_1+p)^2-m_t^2, & D_2 & =(k_2+p)^2-m_t^2,
\nonumber\\
D_3 & =(k_3+p)^2-m_t^2, & D_4 & =(k_3+k_4)^2,\nonumber\\
D_5 & =(k_2+k_4)^2, & D_6 & =(k_1+k_4)^2,
\nonumber\\
D_7 & =k_1^2, & D_8 & =(k_1-k_2)^2,\nonumber\\
D_9 & =(k_2-k_3)^2, & D_{10} & =(k_1-k_3)^2,
\nonumber\\
D_{11} & =k_2^2, & D_{12} & =k_3^2,\nonumber\\
D_{13} & =k_4^2, & D_{14} & =(k_4-p)^2-m_W^2,\nonumber
\label{int}
\end{align}
and
$
{\mathcal D}^d k_i \equiv -i \pi^{-d/2}e^{-\epsilon \gamma} m_t^{2\epsilon} \dd[d]{k_i},~ d=4-2\epsilon \ .
$
The external top quarks admit the on-shell condition $p^2=m_t^2$.
In practice, only the integrals with non-vanishing imaginary parts are relevant, which simplifies the calculation.
For the leading color contribution, all scalar integrals are reduced into 185 MIs. 
These MIs involve two scales, $m_W^2$ and $m_t^2$,
and therefore they can be expressed as functions of a single parameter $w$.
Adopting the differential equation method~\cite{Kotikov:1990kg,Kotikov:1991pm, Henn:2013pwa},  we remarkably succeed in constructing the canonical basis $\vb{B}$ of the MIs which satisfies the differential equation
\bqa
\dv{\vb{B}}{w} = \epsilon \qty( \frac{\vb{P}}{w} + \frac{\vb{N}}{w-1} ) \vb{B} 
\eqa
with $\vb{P}$ and $\vb{N}$ rational matrices. In some low sectors, the package \texttt{Libra} \cite{Lee:2020zfb} has been used to transform coupled integrals to the canonical basis.
A tentative choice of a basis integral with more propagators than the normal ones in a specified sector proves helpful in achieving a compact form of the basis. 

Solving the above equation recursively, we obtain analytical expressions for basis integrals in a series of $\epsilon$ with undetermined constants.
The coefficient of each order of $\epsilon$ has been written in terms of harmonic polylogarithms \cite{Remiddi:1999ew}.
We evaluate these expressions in a fixed value of $w$ within the region $(0,1)$, e.g., $w=1/4$,
using the {\tt HPL} package \cite{Maitre:2005uu}, and compare them with the high-precision numerical results by \texttt{AMFlow} \cite{Liu:2017jxz,Liu:2022chg} to determine the integration constants.
Their analytic form is recovered by making use of the PSLQ algorithm \cite{Ferguson1992,Ferguson:1999aa}.
The expressions of the master integrals are cross-checked with \texttt{AMFlow} at arbitrary values of $w$ in $0<w<1$.
Notice that the imaginary part of each four-loop self-energy diagram has only poles up to $1/\epsilon^3$, in which all the IR divergences have canceled while the UV divergences remain.

As for renormalization, we have to calculate lower-loop diagrams with insertions of counter-terms.
These loop integrals are handled with the same method as used in \cite{Chen:2022wit}, but higher order expressions in $\epsilon$ are needed.
This is easily realized in the method of canonical differential equations.
After renormalization, all the UV divergences cancel out,
which is a strong check of our calculation. As another nontrivial test, the decay rate is vanishing for $w=1$ because no phase space exists.

The analytical results of $Y_A$, $Y_l$ and $Y_{l2}$ in Eq.(\ref{eq:x3}) are compact,
and their complete forms and expansions around $w=1$ have been provided in the supplemental material.
Here, we present the expansion series near the boundary $w=0$.
\begin{widetext}
\begin{align} 
Y_A&= \bigg[\frac{203185}{41472}-\frac{12695\pi^2}{1944}-\frac{4525\zeta (3)}{576}-\frac{1109\pi^4}{25920}+\frac{37\pi^2\zeta (3)}{36}+\frac{1145\zeta (5)}{96}+\frac{47\pi^6}{2835}-\frac{3\zeta (3)^2}{4}\bigg]\nonumber \\
&+w\bigg[-\frac{157939}{2304}+\frac{140863\pi^2}{20736}+\frac{5073\zeta (3)}{64}-\frac{14743\pi^4}{6480}-\frac{169\pi^2\zeta (3)}{72}-\frac{45\zeta (5)}{16}+\frac{3953\pi^6}{22680}-\frac{15\zeta (3)^2}{4}\bigg]\nonumber \\
&+w^2\bigg[\log (w)\bigg(\frac{851099}{27648}-\frac{5875\pi^2}{2304}-\frac{33\zeta (3)}{8}+\frac{\pi^4}{10}\bigg)-\frac{82610233}{331776}+\frac{799511\pi^2}{27648} \nonumber \\
&+\frac{4093\zeta (3)}{32}-\frac{5987\pi^4}{2880}-\frac{91\pi^2\zeta (3)}{16}-\frac{275\zeta (5)}{8}+\frac{347\pi^6}{3024}-\frac{9\zeta (3)^2}{8}\bigg]+\mathcal{O}(w^3),\nonumber\\
Y_l&=\qty[\frac{18209}{20736}+\frac{60025\pi^2}{31104}-\frac{197\zeta (3)}{288}-\frac{14\pi^4}{405}+\frac{5\pi^2\zeta (3)}{36}-\frac{25\zeta (5)}{12}]\nonumber \\
&+w\qty[-\frac{179}{1152}-\frac{3709\pi^2}{2592}-\frac{73\zeta (3)}{6}+\frac{46\pi^4}{405}+\frac{19\pi^2\zeta (3)}{18}+\frac{5\zeta (5)}{2}]\nonumber \\
&+w^2\qty[\log(w)\qty(-\frac{11077}{1152}+\frac{37\pi^2}{96}+\frac{3\zeta (3)}{8})+\frac{49097}{648}-\frac{817\pi^2}{128}-\frac{2651\zeta (3)}{96}+\frac{17\pi^4}{270}+\frac{5\pi^2\zeta (3)}{6}+\frac{25\zeta (5)}{4}]+\mathcal{O}(w^3),\nonumber \\
Y_{l2}&= \bigg[-\frac{695}{2592}-\frac{91\pi^2}{972}+\frac{11\zeta (3)}{36}-\frac{2\pi^4}{405}\bigg]
+w\qty[\frac{245}{144}-\frac{73\pi^2}{648}-\frac{\zeta (3)}{3}] \nonumber \\
&+w^2\qty[\log(w)\qty(\frac{245}{432}-\frac{\pi^2}{72})-\frac{791}{162}+\frac{85\pi^2}{432}+\frac{3\zeta (3)}{4}+\frac{2\pi^4}{135}]+\mathcal{O}(w^3),
\end{align}
\end{widetext}
where $\zeta(n)$ is Riemann zeta function.
We have set the renormalization scale $\mu=m_t$ in the above equations, and the full $\mu$ dependent terms can be easily recovered from the running equation of the strong coupling.
One can see that there is a single logarithm $\log(w)$ starting from $\mathcal{O}(w^2)$.
This comes from the one-loop tadpole integral with a scale of $m_W$ when expanding the four-loop integral in the small $w$ limit with the method of regions.

\begin{table}[]
    \centering
    \begin{tabular}{c|ccccc}
    \hline \hline 
            & $\delta_b^{(i)}$  & $\delta_{ W}^{(i)}$ & $\delta_{\rm EW}^{(i)}$ & $\delta_{\rm QCD}^{(i)}$& $\Gamma_t$ [GeV]   \\
    \hline 
        LO    & -0.273  & -1.544   &  $-$ & $-$    &  1.459  \\
        NLO   & 0.126   & 0.132     &  1.683    & -8.575 &  1.361$^{+0.0091}_{-0.0130}$ \\
        NNLO  & $*$     & 0.030    & $*$  & -2.070 &  1.331$^{+0.0055}_{-0.0051}$ \\
        N$^3$LO  & $*$     & 0.009    & $*$  & -0.667 &  1.321$^{+0.0025}_{-0.0021}$ \\        
    \hline \hline 
    \end{tabular}
    \caption{Top-quark width up to N$^3$LO and  various higher-order corrections  in percentage ($\%$) with respect to the LO width $\Gamma_{t}^{(0)} = 1.486$ GeV. 
    The values of $\delta_b^{(1)}$ and $\delta_{\rm EW}^{(1)}$ are calculated using the formulae in \cite{Bohm:1986rj,Jezabek:1988iv,Denner:1990ns,Denner:1990tx}.
    The notation `$*$' represents the correction that has not been calculated.
    The numbers in the last column show the decay width with all the available corrections up to that order.
   The scale uncertainties are also shown explicitly.}
    \label{tab:my_label}
\end{table}

\begin{figure}
    \centering
    \includegraphics[width=0.95\linewidth]{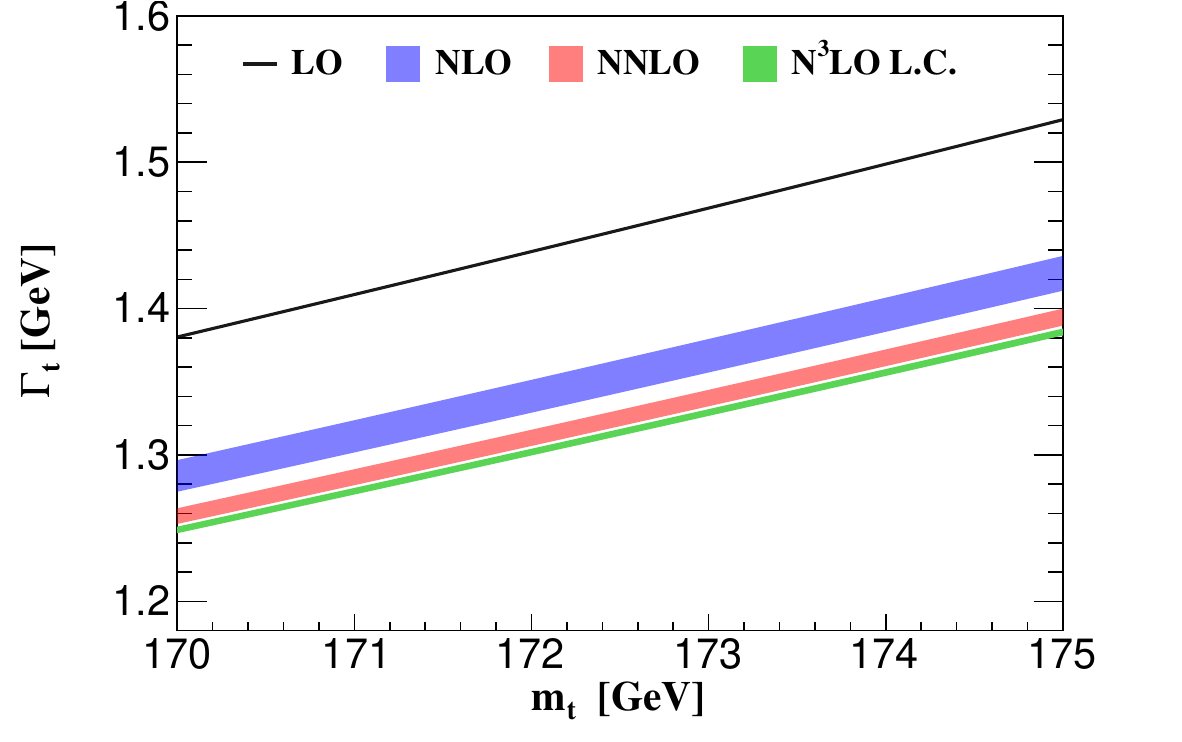}
    \caption{Top-quark width at different values of $m_t$. The black line shows the LO result. The NLO, NNLO, and N$^3$LO predictions with QCD scale uncertainties are represented by the blue, red, and green bands, respectively.}
    \label{fig:scl}
\end{figure}

To see the impact of the ${\rm N^3LO}$ QCD corrections, we decompose the decay width according to the perturbative orders,
\begin{align}
    \Gamma_t & = \Gamma_t^{(0)} [(1+\delta_b^{(0)} + \delta_W^{(0)}) \nn \\
    & \qquad + (\delta_b^{(1)} + \delta_W^{(1)}  +\delta_{\rm EW}^{(1)} +\delta_{\rm QCD}^{(1)}  ) \\ 
    & \qquad + (\delta_b^{(2)} + \delta_W^{(2)}  +\delta_{\rm EW}^{(2)} +\delta_{\rm QCD}^{(2)}  +\delta_{\rm EW \times QCD}^{(2)}   ) \nn \\
    & \qquad + (\delta_b^{(3)} + \delta_W^{(3)}  +\delta_{\rm EW}^{(3)} +\delta_{\rm QCD}^{(3)}  +\delta_{\rm EW \times QCD}^{(3)}   )     
    ],\nn
\end{align}
where the LO width $\Gamma_{t}^{(0)} = 1.486$ GeV is obtained with $m_b=0$ and on-shell $W$.
The corrections from finite $b$ quark mass
effect and off-shell $W$ boson contribution are denoted by $\delta_b^{(i)}$ and $\delta_W^{(i)}$, respectively.
The higher-order QCD and EW corrections are labeled as $\delta_{\rm QCD}^{(i)}$ and $\delta_{\rm EW}^{(i)}$, respectively.
Adopting the SM input parameters \cite{ParticleDataGroup:2022pth}
\begin{align}
    m_t & = 172.69 ~{\rm GeV}, & m_b & = 4.78~{\rm GeV},  \nn \\
    m_W & = 80.377 ~{\rm GeV}, & \Gamma_W & = 2.085~{\rm GeV}, \\
    m_Z & = 91.1876 ~{\rm GeV}, & G_F & = 1.16638 \times 10^{-5}~{\rm GeV}^{-2}, \nn
    \label{eq:input}    
\end{align}
and choosing $|V_{tb}| =1 $  and $\alpha_s(m_Z) = 0.1179$,
the N$^3$LO QCD correction is $-0.667\%$ of the LO result $\Gamma_{t}^{(0)}$.
Adding all the other higher-order corrections that have been discussed in detail in our previous paper \cite{Chen:2022wit}, 
we obtain the most accurate prediction for the top-quark width $\Gamma_t = 1.321$ GeV at $m_t=172.69$ GeV.
The scale uncertainty is reduced to $\pm0.2\%$, only half of that at NNLO.
Now the N$^3$LO and NNLO results with uncertainties are almost adjacent to each other, displaying good convergent behavior.
All the formulas in the calculation are given in analytic form.
The strong coupling $\alpha_s$ at different scales is related by an analytic solution to the three-loop renormalization group evolution equation~\cite{Gardi:1998qr,Deur:2016tte}.
As a result, our calculation is efficient and accurate.
Readers can perform a customized calculation using the 
Mathematica program {\tt TopWidth} 
\footnote{
The program can be downloaded from
 \url{https://github.com/haitaoli1/TopWidth}.
 }.
In Fig.\ref{fig:scl}, we show the top-quark decay width for $170{~\rm GeV}\le m_t \le 175 $ GeV.
A nearly linear dependence can be observed.
For the convenience of readers, we provide a fitted function for the top-quark width within this range,
\begin{align}
    \Gamma_t(m_t) =  0.027037 \times m_t  - 3.34801 ~{\rm GeV}\,.
\end{align}

Our result of the top-quark decay can readily be applied in the calculation of the  $b$-quark semileptonic decay $b\to X_u e\bar{\nu}_e$.
The dilepton invariant mass spectrum up to N$^3$LO is given by
\begin{align}
    \dv{\Gamma(b\to X_u e \bar{\nu}_e) }{q^2} = \Gamma_b^{(0)}
 \sum_{i=0}^3 \left(\frac{\alpha_s}{\pi}\right)^i X_i\left(\frac{q^2}{m_b^2}\right) .
\end{align}
with $\Gamma_b^{(0)}=G_F^2|V_{ub}|^2m_b^3 / 96\pi^3$.
We have replaced the argument of $X_i$ by $q^2/m_b^2$.
The $b$-quark semileptonic decay width can be expanded in $\alpha_s$, 
\begin{align}
\Gamma(b\to X_u e \bar{\nu}_e)  =\frac{G_F^2|V_{ub}|^2m_b^5}{192\pi^3}
 \left[1
 +\sum_{i=1}\left(\frac{\alpha_s}{\pi}\right)^i b_i \right].
\end{align}
The NLO and NNLO coefficients, $b_1$ and $b_2$, have been calculated in \cite{vanRitbergen:1999gs}.
Integrating $X_3$ in Eq.(\ref{eq:x3}) over $w$ in the region $[0,1]$, we obtain the result for $b_3$,
\bqa
b_3&=& C_F\bigg[ N_c^2\bigg(\frac{9651283}{82944}-\frac{1051339 \pi ^2}{62208}-\frac{67189 \zeta (3)}{864}
\nonumber\\
&+&\frac{4363 \pi ^4}{6480}+\frac{59 \pi ^2 \zeta (3)}{32}+\frac{3655 \zeta (5)}{96}-\frac{109 \pi ^6}{3780}\bigg)
\nonumber\\
&+ &n_l N_c \bigg(-\frac{729695}{27648}+\frac{48403 \pi ^2}{15552}+\frac{1373 \zeta (3)}{108}+\frac{133 \pi ^4}{1728}\nonumber\\
&- &\frac{13 \pi ^2 \zeta (3)}{72}-\frac{125 \zeta(5)}{24}\bigg)+n_l^2\bigg(\frac{24763}{20736}-\frac{1417\pi ^2}{15552}\nonumber\\
&-&\frac{37 \zeta (3)}{216}-\frac{121 \pi ^4}{6480}\bigg)+\rm{subleading~color} \bigg]\nn\\
&=& (- 195.3 \pm 9.8) C_F\,.
\eqa
In the last line, we give a numerical estimate of the subleading color contribution, which is about $5\%$ of the leading color result as indicated at NNLO.
Our result is consistent with the estimation in \cite{Fael:2020tow}, $b_3=(-202\pm 20)C_F$, which is obtained by taking the expansion in terms of $\delta=1-m_u/m_b$.

To summarize, we have obtained the first analytic N$^3$LO QCD leading color corrections to the top-quark decay width.  
This is accomplished by applying the optical theorem. 
The imaginary parts of four-loop integrals 
have been calculated with the differential equation method.
All the divergences cancel out after renormalization.
The final result of the decay width is vanishing if setting $m_W=m_t$ 
and exhibits a single logarithmic dependence on $m_W$ when expanded around $m_W=0$.
These features serve as nontrivial checks.
The  N$^3$LO QCD corrections decrease the LO result by $-0.667\%$.
Combining with the other higher-order corrections, such as the EW corrections and off-shell $W$ effects, 
we get the most precise theoretical prediction of the top-quark width, $\Gamma_t=1.321$ GeV at $m_t=172.69$ GeV,
with a scale uncertainty of $\pm0.2\%$.

Furthermore, we derive the analytic third-order QCD leading color predictions for the semileptonic $b\to u$ decay width and the dilepton invariant mass spectrum, which are useful in the precise determination of the CKM matrix element $V_{ub}$.

Our calculation can be extended to more differential observables, such as the invariant mass of hadronic final states or the decay into polarized $W$ bosons.
It is also interesting to understand the simple analytic structure of four-loop integrals with massive propagators in our case.
Especially, we are curious about the explanations of the symbol letters with Landau equations \cite{Dlapa:2023cvx}, intersection theory \cite{Chen:2023kgw}, or some other methods.  

{\it Note added}: When we were finishing this manuscript, we became aware of the result by L. Chen, X. Chen, X. Guan and Y.-Q. Ma \cite{chen:2023xx} which was obtained by numerical calculation with {\tt AMFlow}.
We have compared the leading color results and found perfect agreement.
Their results confirm that the leading color contribution accounts for $95\%$ of the N$^3$LO correction, similar to the case at NNLO.

{\it Acknowledgments}:
This work was supported in part by the National Natural Science Foundation of China under grant  Nos. 12005117, 12075251, 12147154, 12175048, 12275156, 12321005, 12375076.  
The work of L.B.C. was also supported by the Natural Science Foundation of Guangdong Province under grant No. 2022A1515010041. The work of J.W. was also supported by the Taishan Scholar Project of Shandong Province (tsqn201909011).

\bibliographystyle{apsrev}
\bibliography{reference}

\onecolumngrid
\newpage

\section*{Supplemental Material}

The complete analytic results of $Y_A, Y_l$ and $Y_{l2}$ are presented below.
\begin{align}
Y_A &= \frac{1}{8}\Bigg((2 w+1) (w-1)^2 \Big(7 \zeta (3) H_{1,0,1}(w)+\frac{4}{15} \pi ^4 H_{1,0}(w)+\frac{1}{6} \pi ^2 H_{1,0,0,1}(w)-\frac{2}{3} \pi ^2 H_{1,0,1,0}(w)+9 H_{1,0,0,0,1,1}(w)
\nonumber\\ &
-H_{1,0,0,1,0,1}(w)-4 H_{1,0,0,1,1,0}(w)-5 H_{1,0,1,0,0,1}(w)+H_{1,0,1,0,1,0}(w)-\left(2 \pi ^2 \zeta (3)+5 \zeta (5)\right) H_1(w)\Big)
\nonumber\\ &
+\frac{1}{3} \pi ^2 w \left(2 w^2-9 w-10\right) H_{0,1,1,0}(w)+4 (2 w+1) \left(w^2-2 w+4\right) H_{0,0,1,0,1,1}(w)+6 \left(6 w^2+6 w-1\right) H_{0,1,0,1,1,0}(w)
\nonumber\\ &
+\left(18 w^3+15 w^2+82 w+22\right) \zeta (3) H_{0,0,1}(w)+4 \left(4 w^3+24 w^2+50 w+7\right) \zeta (3) H_{0,1,1}(w)
\nonumber\\ &
-\frac{1}{45} \pi ^4 \left(130 w^3+201 w^2+612 w+107\right) H_{0,1}(w)-\frac{1}{6} \pi ^2 \left(2 w^3-93 w^2-54 w+34\right) H_{0,0,0,1}(w)
\nonumber\\ &
-\frac{4}{3} \pi ^2 \left(4 w^3+9 w^2+22 w+3\right) H_{0,0,1,0}(w)-\frac{2}{3} \pi ^2 \left(14 w^3+3 w^2+58 w+20\right) H_{0,0,1,1}(w)
\nonumber\\ &
+\frac{2}{3} \pi ^2 \left(4 w^3+18 w^2+22 w-3\right) H_{0,1,0,1}(w)-\frac{2}{3} \pi ^2 \left(2 w^3+3 w^2+10 w+2\right) H_{0,1,1,1}(w)
\nonumber\\ &
+\left(38 w^3+33 w^2+174 w+46\right) H_{0,0,0,0,1,1}(w)+\left(-6 w^3+3 w^2-22 w-10\right) H_{0,0,0,1,0,1}(w)
\nonumber\\ &
-4 \left(4 w^3-3 w^2+14 w+7\right) H_{0,0,0,1,1,0}(w)+\left(-18 w^3-51 w^2-106 w-10\right) H_{0,0,1,0,0,1}(w)
\nonumber\\ &
+2 \left(2 w^3+57 w^2+100 w+11\right) \left(2 H_{0,0,0,1,1,1}(w)-H_{0,0,1,1,0,1}(w)\right)-2 \left(6 w^3+51 w^2+112 w+19\right) H_{0,0,1,1,1,0}(w)
\nonumber\\ &
+4 \left(2 w^3+6 w^2+12 w+1\right) \left(-H_{0,0,0,0,0,1}(w)+H_{0,0,0,0,1,0}(w)-H_{0,1,0,0,0,1}(w)+H_{0,1,0,0,1,0}(w)\right)
\nonumber\\ &
-2 \left(2 w^3+51 w^2+78 w+4\right) H_{0,1,0,0,1,1}(w)-2 \left(4 w^3+12 w^2+30 w+5\right) H_{0,1,0,1,0,1}(w)
\nonumber\\ &
+4 \left(2 w^3+9 w^2+20 w+3\right) H_{0,1,1,0,0,1}(w)+2 \left(2 w^3+27 w^2+50 w+6\right) H_{0,1,1,0,1,0}(w)
\nonumber\\ &
+\left(2 w^3+3 w^2+10 w+2\right) \left(H_{0,0,1,0,1,0}(w)-4 H_{0,1,0,1,1,1}(w)+4 H_{0,1,1,1,0,1}(w)\right)
\nonumber\\ &
-\frac{\pi^6\left(538w^3-5205w^2-7906w-752\right)}{5670}
-3\left(2w^3+3w^2+10w+2\right)\zeta(3)^2\Bigg)\nonumber\\
 &-\frac{11}{4} (2 w+1)(w-1)^2 \zeta (3) H_{1,0}(w) +\frac{1}{60} \pi ^4 w \left(2 w^2+5 w-2\right) H_0(w)\nonumber\\
 &+\frac{1}{720} \pi ^4 \left(940 w^3+27w^2-918 w-49\right) H_1(w)+\frac{1}{96} \left(2416 w^3+468 w^2+10026 w+2987\right) \zeta (3) H_{0,1}(w)\nonumber\\
 &+\frac{1}{4} \left(36 w^3-165 w^2+52
   w+77\right) \zeta (3) H_{1,1}(w)+\frac{1}{576} \pi ^2 \left(88 w^3-720 w^2-5046 w-2167\right) H_{0,0,1}(w)\nonumber\\
 &+\frac{1}{96} \pi ^2 \left(92
   w^3+374 w^2+162 w-29\right) H_{0,1,0}(w)+\frac{1}{72} \pi ^2 \left(392 w^3+105 w^2+632 w+150\right) H_{0,1,1}(w)\nonumber\\
 &-\frac{1}{144} \pi ^2
   \left(82 w^3+93 w^2-264 w+89\right) H_{1,0,1}(w)-\frac{1}{72} \pi ^2 \left(94 w^3-168 w^2+3 w+71\right) H_{1,1,0}(w)\nonumber\\
 &+\frac{1}{12} \pi ^2
   \left(2 w^3+9 w^2-7 w-4\right) H_{1,1,1}(w)-\frac{1}{48} \left(146 w^3-27 w^2+688 w+149\right) H_{0,0,0,0,1}(w)\nonumber\\
 &+\frac{1}{48} \left(146
   w^3-27 w^2+688 w+149\right) H_{0,0,0,1,0}(w)-\frac{1}{96} \left(224 w^3+2988 w^2+34 w-943\right) H_{0,0,0,1,1}(w)\nonumber\\
 &-\frac{1}{96} \left(40
   w^3-984 w^2-982 w+67\right) H_{0,0,1,0,1}(w)+\frac{1}{48} \left(128 w^3+1008 w^2+656 w-231\right) H_{0,0,1,1,0}(w)\nonumber\\
 &+\frac{1}{24} \left(108
   w^3-1848 w^2-1162 w+407\right) H_{0,0,1,1,1}(w)-\frac{1}{96} \left(168 w^3-648 w^2+2146 w+601\right) H_{0,1,0,0,1}(w)\nonumber\\
&+\frac{1}{96} \left(176
   w^3-660 w^2-114 w+187\right) H_{0,1,0,1,0}(w)-\frac{1}{24} \left(156 w^3+444 w^2+866 w+89\right) H_{0,1,0,1,1}(w)\nonumber\\
&+\frac{1}{24} \left(56
   w^3+1167 w^2+1250 w-75\right) H_{0,1,1,0,1}(w)-\frac{1}{24} \left(8 w^3-1125 w^2-778 w+243\right) H_{0,1,1,1,0}(w)\nonumber\\
&-\frac{1}{48} \left(14
   w^3-327 w^2+216 w+97\right) H_{1,0,0,0,1}(w)+\frac{1}{48} \left(14 w^3-327 w^2+216 w+97\right) H_{1,0,0,1,0}(w)\nonumber\\
&+\frac{1}{24} \left(82
   w^3+345 w^2-234 w-193\right) H_{1,0,0,1,1}(w)-\frac{1}{24} \left(20 w^3-237 w^2+108 w+109\right) H_{1,0,1,0,1}(w)\nonumber\\
&-\frac{1}{12} \left(13
   w^3+21 w^2-36 w+2\right) H_{1,0,1,1,0}(w)+\frac{1}{2} \left(2 w^3+9 w^2-7 w-4\right) H_{1,0,1,1,1}(w)\nonumber\\
&-\frac{1}{12} \left(56 w^3+51 w^2-66
   w-41\right) H_{1,1,0,0,1}(w)+\frac{1}{12} \left(38 w^3-219 w^2+69 w+112\right) H_{1,1,0,1,0}(w)\nonumber\\
&-\frac{1}{2} \left(2 w^3+9 w^2-7 w-4\right)
   H_{1,1,1,0,1}(w)+\frac{5}{96} \left(344 w^3-654 w^2-42 w+229\right) \zeta (5)\nonumber\\
&+\frac{1}{144} \pi ^2 \left(186 w^3-801 w^2-302 w+148\right)
   \zeta (3)
+\frac{\pi ^2 \left(3898 w^4-1817 w^3+7480 w^2-4503 w-90\right) H_{0,1}(w)}{1152 w}\nonumber\\&-\frac{\pi ^2 \left(14978 w^4-8265 w^3-11340 w^2+4411 w+216\right) H_{1,0}(w)}{3456 w}\nonumber\\&
-\frac{\pi ^2 \left(6510 w^4-1833 w^3-6664 w^2+1951 w+36\right) H_{1,1}(w)}{576 w}\nonumber\\&-\frac{\left(3582
   w^4-16281 w^3+11248 w^2+6457 w+36\right) (H_{0,0,0,1}(w)- H_{0,0,1,0}(w))}{576 w}\nonumber\\&+\frac{\left(1578 w^4+4131 w^3-22606 w^2-1816 w+153\right) H_{0,0,1,1}(w)}{288 w}\nonumber\\&-\frac{\left(1906 w^4-3744 w^3-3168
   w^2+473 w+27\right) H_{0,1,0,1}(w)}{288 w}\nonumber\\&-\frac{\left(530 w^4+6537 w^3-32312 w^2-3043 w+180\right) H_{0,1,1,0}(w)}{576 w}\nonumber\\&-\frac{\left(4610
   w^4-29343 w^3+11820 w^2+12805 w+108\right) H_{0,1,1,1}(w)}{288 w}\nonumber\\&-\frac{\left(8818 w^4-13227 w^3-2928 w^2+7211 w+126\right)
   H_{1,0,0,1}(w)}{576 w}\nonumber\\&+\frac{\left(5002 w^4-11220 w^3+1818 w^2+4373 w+27\right) H_{1,0,1,0}(w)}{288 w}+\frac{\left(153 w^4+936 w^3-773
   w^2-319 w+3\right) H_{1,0,1,1}(w)}{24 w}\nonumber\\&-\frac{\left(380 w^4+6807 w^3-4112 w^2-3057 w-18\right) H_{1,1,0,1}(w)}{96 w}\nonumber\\&
   +\frac{\left(1957
   w^4-10077 w^3+4380 w^2+3731 w+9\right) H_{1,1,1,0}(w)}{144 w}\nonumber\\&-\frac{11}{4} w \left(2 w^2+w-1\right) \zeta (3) H_0(w)-\frac{\left(14786
   w^4+64419 w^3-48276 w^2-30695 w-234\right) \zeta (3) H_1(w)}{576 w}\nonumber\\&+\frac{\pi ^4 \left(33778 w^3+26253 w^2-48640 w-1109\right)}{25920}+\frac{\left(2237 w^2-4622 w+222\right) (w-1)^2 H_{1,1,1}(w)}{288 w}\nonumber\\
&-\frac{\left(65200 w^4-110430 w^3+2268 w^2+76589 w+1368\right)
   H_{0,0,1}(w)}{3456 w}\nonumber\\
&+\frac{\left(85618 w^4-53034 w^3+11634 w^2+77021 w+1368\right) H_{0,1,0}(w)}{3456 w}\nonumber\\
&+\frac{\left(-117644 w^4+720114
   w^3+459660 w^2-170179 w+1692\right) H_{0,1,1}(w)}{3456 w}\nonumber\\
&+\frac{\left(1858 w^4-27381 w^3+12990 w^2+12518 w+15\right) H_{1,0,1}(w)}{288w}\nonumber\\
&+\frac{\left(71305 w^4-221919 w^3+23019 w^2+127829 w-234\right) H_{1,1,0}(w)}{1728 w}
\nonumber\\&
+\frac{\pi ^2 \left(-12119 w^3-8710 w^2+7291
   w+216\right) H_0(w)}{3456}\nonumber\\
&-\frac{\pi ^2 \left(88195 w^4-66021 w^3-59835 w^2+36521 w+1140\right) H_1(w)}{3456 w}-\frac{1}{576} \left(23773
   w^3+67462 w^2+6245 w+4759\right) \zeta (3)\nonumber\\
&-\frac{\left(267850 w^4+1706859 w^3-1838292 w^2+631646 w+65376\right) H_{0,1}(w)}{41472 w}\nonumber\\
&+\frac{\left(938906 w^4-3991935 w^3+1309482
   w^2+1710859 w+32688\right) H_{1,0}(w)}{20736 w}\nonumber\\
&+\frac{\left(571502 w^5-3511083 w^4+1727700 w^3+1217512 w^2-5874 w+243\right)H_{1,1}(w)}{13824 w^2}\nonumber\\
&+\frac{\pi ^2 \left(-7155530 w^3+3914433 w^2+5613528 w-1538992\right)}{248832}+\frac{4154412 w^3-7897452 w^2+2657821 w+1085219}{82944}\nonumber\\
&+\frac{\left(5254190 w^3-14205413 w^2-9986716 w-168768\right) H_{0}(w)}{82944}\nonumber\\
&+\frac{\left(6497054 w^4-22591845 w^3+5613132 w^2+10780501 w-298842\right) H_{1}(w)}{82944 w},\nonumber
\end{align}
\bqa
Y_l &=& \frac{1}{12}\Bigg((2w+1)(w-1)^2 \Big(-3 \zeta (3) \left(4 H_{1,1}(w)-H_{1,0}(w)\right)+\frac{2}{3} \pi ^2 \left(H_{1,0,1}(w)+2 H_{1,1,0}(w)\right)-2 H_{0,1,0,1,0}(w)
\nonumber\\& &
+H_{1,0,0,0,1}(w)-H_{1,0,0,1,0}(w)-4 H_{1,0,0,1,1}(w)+2 H_{1,0,1,0,1}(w)+2 H_{1,0,1,1,0}(w)+4 H_{1,1,0,0,1}(w)-4 H_{1,1,0,1,0}(w)
\nonumber\\& &
-\frac{16}{15}\pi ^4 H_1(w)\Big)-\frac{1}{3} \pi ^2 \left(28 w^3+64 w+21\right) H_{0,1,1}(w)-\left(50 w^3+33 w^2+198 w+43\right) \zeta (3) H_{0,1}(w)
\nonumber\\& &
+\frac{1}{3} \pi ^2 \left(2 w^3-21 w^2-18 w-2\right) H_{0,0,1}(w)-\frac{1}{2} \pi ^2 \left(2 w^3+9 w^2+18 w+3\right) H_{0,1,0}(w)
\nonumber\\& &
+\left(8 w^3+12 w^2+43 w+8\right) \left(H_{0,0,0,0,1}(w)-H_{0,0,0,1,0}(w)\right)+\left(-2 w^3+51 w^2+86 w+7\right) H_{0,0,0,1,1}(w)
\nonumber\\& &
+\left(2 w^3-15 w^2-20 w-1\right) H_{0,0,1,0,1}(w)+\left(-2 w^3-45 w^2-80 w-9\right) H_{0,0,1,1,0}(w)
\nonumber\\& &
-2 \left(6 w^3-33 w^2-46 w-1\right) H_{0,0,1,1,1}(w)+\left(6 w^3+3 w^2+14 w+5\right) H_{0,1,0,0,1}(w)
\nonumber\\& &
+2 \left(6 w^3+21 w^2+50 w+8\right) H_{0,1,0,1,1}(w)-2 \left(2 w^3+27 w^2+56 w+6\right) H_{0,1,1,0,1}(w)
\nonumber\\& &
+2 \left(2 w^3-27 w^2-40 w-3\right) H_{0,1,1,1,0}(w)+\frac{1}{3} \pi ^2 \left(30 w^2+38 w+5\right) \zeta (3)
\nonumber\\& &
+5 \left(-10 w^3+15 w^2+6 w-5\right) \zeta (5)\Bigg)\nonumber\\
& & + \frac{1}{864} \pi ^2 \left(182 w^3+147 w^2-1848 w-191\right) H_{0,1}(w)+\frac{1}{96} \pi ^2 \left(70 w^3-39 w^2-40 w+9\right)
H_{1,0}(w)\nonumber\\
& &+\frac{1}{144} \pi ^2 \left(458 w^3-345 w^2-268 w+155\right) H_{1,1}(w)+\frac{1}{144} \left(190 w^3-825 w^2+622 w+354\right)
H_{0,0,0,1}(w)\nonumber\\
& &-\frac{1}{144} \left(190 w^3-825 w^2+622 w+354\right) H_{0,0,1,0}(w)+\frac{1}{72} \left(70 w^3-420 w^2+424 w+339\right)
H_{0,0,1,1}(w)\nonumber\\
& &+\frac{1}{36} \left(32 w^3-27 w^2-9 w-8\right) H_{0,1,0,1}(w)-\frac{1}{144} \left(86 w^3-729 w^2+908 w+513\right)
H_{0,1,1,0}(w)\nonumber\\
& &+\frac{1}{72} \left(406 w^3-1167 w^2+120 w+641\right) H_{0,1,1,1}(w)+\frac{1}{48} \left(166 w^3-315 w^2+56 w+93\right)
H_{1,0,0,1}(w)\nonumber\\
& &-\frac{1}{36} \left(170 w^3-291 w^2+18 w+103\right) H_{1,0,1,0}(w)+\left(-w^3-8 w^2+\frac{29 w}{6}+\frac{25}{6}\right)
H_{1,0,1,1}(w)\nonumber\\
& &+\frac{1}{24} \left(14 w^3+231 w^2-104 w-141\right) H_{1,1,0,1}(w)-\frac{1}{36} \left(188 w^3-525 w^2+78 w+259\right)
H_{1,1,1,0}(w)\nonumber\\
& &+\frac{1}{4} w \left(2 w^2+w-1\right) \zeta (3) H_0(w)+\frac{1}{144} \left(1462 w^3+1065 w^2-1872 w-655\right) \zeta (3)
H_1(w)\nonumber\\
& &+\frac{\pi ^4 \left(-225 w^3+174 w^2+328 w-56\right)}{1620}-\frac{\left(265 w^2+212 w+6\right) (w-1)^2 H_{1,1,1}(w)}{72 w}\nonumber\\
& & +\frac{\left(5102 w^4-7917 w^3+522 w^2+3718 w+72\right) H_{0,0,1}(w)}{864
   w}-\frac{\left(3295 w^4-2700 w^3-30 w^2+1859 w+36\right) H_{0,1,0}(w)}{432 w}\nonumber\\
& & +\frac{\left(1750 w^4-8913 w^3-2421 w^2+2579 w-18\right)
   H_{0,1,1}(w)}{216 w}-\frac{\left(190 w^4-4413 w^3+1308 w^2+2927 w-12\right) H_{1,0,1}(w)}{288 w}\nonumber\\
& & +\frac{\left(-5549 w^4+12189 w^3+3 w^2-6661 w+18\right) H_{1,1,0}(w)}{432 w}+\frac{1}{576} \pi ^2 w \left(416 w^2+853 w+18\right) H_0(w)\nonumber\\
& & +\frac{\pi ^2 \left(3876 w^4-4285 w^3-810
   w^2+1203 w+16\right) H_1(w)}{576 w}+\frac{1}{288} \left(5254 w^3+1618 w^2-1090 w-197\right) \zeta (3)\nonumber\\
& &+\frac{\left(688 w^4-48708 w^3-11202 w^2+11723 w+2052\right) H_{0,1}(w)}{5184 w}\nonumber\\
& &-\frac{\left(34016 w^4-119253 w^3+41652 w^2+42559 w+1026\right)
   H_{1,0}(w)}{2592 w}\nonumber\\
& &-\frac{\left(3004 w^4-9123 w^3+2085 w^2+4160 w-126\right) H_{1,1}(w)}{216 w}+\frac{\pi ^2 \left(266324 w^3-173880
   w^2-106158 w+59161\right)}{31104}\nonumber\\
& &+\frac{\left(-194684 w^3+355103 w^2+249304 w+4968\right) H_0(w)}{10368}
\nonumber\\
& & -\frac{\left(125638 w^4-326034 w^3+54897 w^2+151757 w-6258\right)
   H_1(w)}{5184 w}\nonumber\\
& & -\frac{62863 w^3}{3456}+\frac{250849 w^2}{6912}-\frac{173473 w}{10368}-\frac{28423}{20736},\nonumber\\
Y_{l2} &=& -\frac{(w-1)^2 (2 w+1)}{1620}\Bigg(180 \pi ^2 H_{1,1}(w)+90 H_{0,0,0,1}(w)-90 H_{0,0,1,0}(w)+180 H_{0,0,1,1}(w)-90 H_{0,1,1,0}(w)
\nonumber\\& &
+360 H_{0,1,1,1}(w)+180H_{1,0,0,1}(w)-270 H_{1,0,1,0}(w)-360 H_{1,1,1,0}(w)+720 H_1(w) \zeta (3)+8 \pi ^4+75 \pi ^2 H_{0,1}(w)
\nonumber\\& &
+15 \pi ^2 H_{1,0}(w)\Bigg)+\frac{1}{54} \left(25 w^3-30 w^2+6 w+5\right) H_{0,1,0}(w)-\frac{1}{27} \left(7 w^3-39 w^2+15 w+5\right) H_{0,1,1}(w)\nonumber\\& &-\frac{1}{2}
\left(w^2-1\right) H_{1,0,1}(w)-\frac{1}{54} (19 w+5) (w-1)^2 H_{0,0,1}(w)+\frac{1}{108} (88 w+35) (w-1)^2 H_{1,1,0}(w)\nonumber\\& &+\frac{1}{9} (4 w+5)
(w-1)^2 H_{1,1,1}(w)-\frac{1}{108} \pi ^2 w \left(2 w^2+w-1\right) H_0(w)-\frac{1}{216} \pi ^2 \left(80 w^3-153 w^2+30 w+43\right)
H_1(w)\nonumber\\& &+\frac{1}{36} \left(-58 w^3+35 w^2+4 w+11\right) \zeta (3)+\frac{\left(115 w^4+324 w^3-456 w^2+11 w-9\right) H_{0,1}(w)}{324 w}\nonumber\\& &+\frac{\left(230 w^4-642 w^3+216 w^2+187 w+9\right) H_{1,0}(w)}{324
w}+\frac{1}{108} \left(124 w^3-159 w^2-108 w-\frac{6}{w}+149\right) H_{1,1}(w)\nonumber\\& &+\frac{\pi ^2 \left(-1015 w^3+972 w^2+240 w-182\right)}{1944}+\frac{1}{648} \left(709 w^3-517 w^2-479 w-18\right) H_0(w)\nonumber\\& &+\frac{\left(2012 w^4-3141 w^3-384 w^2+1627 w-114\right) H_1(w)}{1296 w}+\frac{151 w^3}{108}-\frac{2461 w^2}{864}+\frac{2041 w}{1296}-\frac{323}{2592}.
\eqa

The expansion of the above results around $w=1$ is given by
\begin{align}
Y_A &=(1-w)^2\bigg[\frac{1603421}{4608}-\frac{173167 \pi ^2}{2592}-\frac{37943\zeta (3)}{192}+\frac{5591 \pi ^4}{1728}+\frac{173 \pi ^2 \zeta (3)}{16}+\frac{1465 \zeta (5)}{32}-\frac{6941 \pi ^6}{60480}+\frac{9 \zeta (3)^2}{16}
\nonumber \\ &
+\log(1-w)\left(-\frac{625609}{2304}+\frac{77845 \pi ^2}{1728}+\frac{1451 \zeta (3)}{16}-\frac{379 \pi ^4}{180}\right)+\log^2(1-w)\left(\frac{1229}{16}-\frac{775 \pi ^2}{72}\right)-\frac{1643\log^3(1-w)}{192}\bigg]
\nonumber \\ &
+(1-w)^3\bigg[-\frac{1774211}{31104}+\frac{423497 \pi ^2}{15552}+\frac{5387 \zeta (3)}{48}-\frac{1957 \pi ^4}{1296}-\frac{443 \pi ^2 \zeta (3)}{72}-\frac{1565 \zeta (5)}{48}+\frac{6941 \pi ^6}{90720}-\frac{3 \zeta (3)^2}{8}
\nonumber \\ &
+\log(1-w)\left(-\frac{44999}{5184}-\frac{4399 \pi ^2}{288}-\frac{865 \zeta (3)}{24}+\frac{9 \pi ^4}{10}\right)+\log^2(1-w)\left(\frac{36899}{1728}+\frac{233 \pi ^2}{108}\right)-\frac{4145\log ^3(1-w)}{864}
\bigg]
\nonumber \\ &
+(1-w)^4\bigg[-\frac{12535819}{248832}+\frac{556661 \pi ^2}{124416}-\frac{6709 \zeta (3)}{2304}-\frac{7423 \pi ^4}{34560}+\frac{7 \pi ^2 \zeta (3)}{72}+\frac{25 \zeta (5)}{96}\nonumber \\ &
+\log(1-w)\left(\frac{630541}{20736}-\frac{7553 \pi ^2}{2592}-\frac{71 \zeta (3)}{96}+\frac{11 \pi ^4}{108}\right)
+\log^2(1-w)\left(\frac{263 \pi ^2}{864}-\frac{6463}{6912}\right)-\frac{169 \log ^3(1-w)}{108}
\bigg]\nonumber \\ &
+\mathcal{O}(1-w)^5\,,\nonumber\\
Y_l &= (1-w)^2\bigg[-\frac{273217}{2304}+\frac{92917 \pi ^2}{5184}+\frac{2611 \zeta (3)}{48}-\frac{971 \pi ^4}{4320}-\pi ^2 \zeta (3)-\frac{35 \zeta (5)}{4} 
\nonumber \\ &
+\log(1-w)\left(\frac{101923}{1152}-\frac{1313 \pi ^2}{108}-\frac{103 \zeta (3)}{4}+\frac{29 \pi ^4}{90}\right)+\log ^2(1-w)\left(\frac{28 \pi ^2}{9}-\frac{373}{16}\right)+\frac{115 \log ^3(1-w)}{48}
\bigg]
\nonumber \\ &
+(1-w)^3\bigg[\frac{704789}{15552}-\frac{9961 \pi ^2}{864}-\frac{639 \zeta (3)}{16}+\frac{1027 \pi ^4}{6480}+\frac{2 \pi ^2 \zeta (3)}{3}+\frac{35 \zeta (5)}{6}
\nonumber \\ &
+\log(1-w)\left(-\frac{87719}{5184}+\frac{4909 \pi ^2}{648}+\frac{29 \zeta (3)}{2}-\frac{29 \pi ^4}{135}\right)
+\log^2(1-w)\left(-\frac{625}{432}-\frac{4 \pi ^2}{3}\right)+\frac{205 \log ^3(1-w)}{216}
\bigg]
\nonumber \\ &
+(1-w)^4\bigg[
\frac{5139869}{248832}-\frac{899 \pi ^2}{1944}+\frac{605 \zeta (3)}{576}-\frac{\pi ^4}{405}
+\log(1-w)\left(\frac{5 \zeta (3)}{24}+\frac{323 \pi ^2}{648}-\frac{334421}{20736}\right)
\nonumber \\ &
+\log^2(1-w)\left(\frac{11035}{3456}-\frac{13 \pi ^2}{108}\right)+\frac{49 \log ^3(1-w)}{864}
\bigg]+\mathcal{O}(1-w)^5\,,\nonumber\\
Y_{l2} &= (1-w)^2\bigg[\frac{851}{96}-\frac{281 \pi ^2}{324}-\frac{19 \zeta (3)}{6}-\frac{2 \pi ^4}{45}+\log(1-w)\left(-\frac{907}{144}+\frac{73 \pi ^2}{108}+2 \zeta (3)\right)
\nonumber \\ &
+\log ^2(1-w)\left(\frac{13}{8}-\frac{2 \pi ^2}{9}\right)-\frac{\log ^3(1-w)}{6}\bigg]
\nonumber \\ &
+(1-w)^3\bigg[-\frac{20545}{3888}+\frac{775 \pi ^2}{972}+3 \zeta (3)+\frac{4 \pi ^4}{135}+\log(1-w)\left(\frac{1769}{648}-\frac{109 \pi ^2}{162}-\frac{4 \zeta (3)}{3}\right)
\nonumber \\ &
+\log ^2(1-w)\left(\frac{4 \pi ^2}{27}-\frac{31}{108}\right)-\frac{\log ^3(1-w)}{27}\bigg]
\nonumber \\ &
+(1-w)^4\bigg[-\frac{18271}{15552}-\frac{7 \pi ^2}{81}-\frac{\zeta (3)}{9}
+\log(1-w)\left(\frac{1549}{1296}+\frac{\pi ^2}{36}\right)-\frac{37 \log ^2(1-w)}{108}+\frac{\log ^3(1-w)}{54}\bigg]\nonumber \\ &
+\mathcal{O}(1-w)^5\,.
\end{align}
\label{sec:appendix}

\end{document}